**Real-Space Observation of Ligand Hole State in Cubic Perovskite SrFeO$_3$**

*Shunsuke Kitou,\* Masaki Gen, Yuiga Nakamura, Kunihisa Sugimoto, Yusuke Tokunaga, Shintaro Ishiwata, and Taka-hisa Arima*

S. Kitou, M. Gen, Y. Tokunaga, T.-h. Arima
Department of Advanced Materials Science
The University of Tokyo
Kashiwa 277-8561, Japan
E-mail: kitou@edu.k.u-tokyo.ac.jp

S. Kitou, M. Gen, T.-h. Arima
Center for Emergent Matter Science
RIKEN
Wako 351-0198, Japan

Y. Nakamura
Japan Synchrotron Radiation Research Institute (JASRI)
SPring-8
Hyogo 679-5198, Japan

K. Sugimoto
Department of Chemistry
Kindai University
Osaka 577-8502 Japan

S. Ishiwata
Division of Materials Physics Graduate School of Engineering Science
Osaka University, Toyonaka
Osaka 560-8531, Japan

**Abstract**

An anomalously high valence state sometimes shows up in transition-metal oxide compounds. In such systems, holes tend to occupy mainly the ligand $p$ orbitals, giving rise to interesting physical properties such as superconductivity in cuprates and rich magnetic phases in ferrates. However, no one has ever observed the distribution of ligand holes in real space. Here, we report a successful observation of the spatial distribution of valence electrons in cubic perovskite SrFeO$_3$ by high-energy X-ray diffraction experiments and precise electron density analysis using a core differential Fourier synthesis method. A real-space picture of ligand holes formed by the orbital hybridization of Fe $3d$ and O $2p$ is revealed. The anomalous valence state in Fe is attributed to the considerable contribution of the ligand hole, which is related to the metallic nature and the absence of Jahn-Teller distortions in this system.

**1. Introduction**

The ionic-covalent character of the metal-oxygen bonding in metal oxides has long been discussed. Since the oxygen and metal atomic orbitals mainly contribute to the bonding and antibonding orbitals, respectively, the bonding has a strong ionic nature and hence the formal ionic valence of metal can be well defined in general. Nonetheless, when the formal valence of metal is anomalously high, the contribution of the oxygen $2p$ orbital to the antibonding orbital becomes dominant, presumably resulting in the ligand hole state[1]. The ligand holes are considered to play a significant role in the electrical transport[2],[3] and magnetism[4] in cuprates,[5],[6] nickelates,[7],[8] cobaltates,[9]-[11] and ferrates.[11]-[15]

SrFeO$_3$ is an archetypal tetravalent ferrate compound, in which, at formally, four electrons occupy the Fe $3d$ orbitals. It forms the perovskite-type structure with the cubic space group $Pm\bar{3}m$ (Figure 1a) and exhibits metallic conductivity.[16] The local symmetry at the Fe site is $m\bar{3}m$. Each Fe atom is surrounded by six O atoms to form a regular octahedron without Jahn-Teller distortion, where the $3d$ orbitals are split into the lower-lying triplet ($t_{2g}$) and the higher-lying doublet ($e_g$). The high-spin state of $3d^4$ corresponds to the $t_{2g}^3 e_g^1$ electron configuration, which causes some anisotropy in the valence electron density. However, previous X-ray photoelectron spectroscopy and X-ray absorption spectroscopy measurements suggest that the ground state consists of mixed $3d^4$ and $3d^5\underline{L}$ ($\underline{L}$: ligand hole) states.[11],[12] In the extreme limit of the $3d^5\underline{L}$ state, the electron density around the Fe site should be spherical because of the $t_{2g}^3 e_g^2$ electron configuration. Theoretically, the ligand hole formed by the $2p$-$3d$ hybridization is expected to stabilize novel itinerant magnetic phases.[17],[18] In fact, despite its simple crystal structure, various magnetic phases including a quadruple-$q$ topological spin structure appear in the magnetic-field-versus-temperature phase diagram in SrFeO$_3$.[14],[15] Furthermore, the crystal structure and physical properties are greatly changed by slight oxygen vacancies,[19]-[24] because this system has charge instability of unusual tetravalent iron cations.

Although the ligand holes in the crystal have been observed by spectroscopy measurements,[11],[12] no one has ever seen where the ligand holes exist in real space. To observe the spatial distribution of the holes, the measurement with high-wavevector ($Q$) resolution is indispensable. In this study, we observe the valence electron density distribution of SrFeO$_3$ by electron density analysis using state-of-the-art synchrotron X-ray diffraction (XRD). The number of 3$d$ electrons is estimated from the anisotropic distribution of the valence electron density around the Fe site. It is also confirmed that the valence electron density around the O site along the Fe—O—Fe axis is slightly reduced.

## 2. Results and Discussion

### 2.1. Crystal structure of SrFeO$_3$

The XRD experiments of SrFeO$_3$ detected no structural phase transitions down to 30 K, which is consistent with the previous neutron diffraction experiment.[23] As a result of the structural analysis, no signs of oxygen vacancy and no anomaly in the atomic displacement parameters of oxygen were confirmed. Here, the structural parameters were determined with high accuracy by performing a high-angle analysis utilizing the advantages of high-energy X-rays,[25] where only reflections with $\sin\theta/\lambda > 0.6$ Å$^{-1}$ ($d < 0.833$ Å) were used. The obtained structural parameters of SrFeO$_3$ are summarized in Tables S1 and S2 in Supporting Information.

### 2.2. Valence electron density around the Fe site

Figure 1b shows the valence electron density distribution of SrFeO$_3$ at 30 K. No valence electron density larger than $3e/$Å$^3$ is observed around the Sr site, which is consistent with the Sr$^{2+}$ ($5s^0$) state. In contrast, valence electrons are observed around the Fe and O sites, as shown by yellow iso-density surfaces. An orange iso-density surface for higher electron distribution is observed only around the Fe site.

Figure 2a shows the iso-density surface around the Fe site with the site symmetry $m\bar{3}m$. The shape is clearly distinct from a sphere: there are six hollow holes toward the six ligand oxygens. To quantify the anisotropy of the valence electron density $\rho(\boldsymbol{r})$ around the Fe site, the density at a distance $r = 0.2$ Å from the Fe nucleus, which corresponds to the peak top of $\rho(r)$ (see Figure 3), is shown by a color map on a sphere (Figure 2b). The maximum and minimum electron densities are present along the <111> and <100> axes, respectively; $\rho_{\max} =$

$10.76 e/Å^3$ and $\rho_{\min} = 9.85 e/Å^3$. Figures 2c and 2d show surface color maps of $\rho(\theta, \phi)$ for the calculated electron density considering the high-spin $3d^4$ and $3d^5$ states for an isolated Fe ion, respectively. To accurately evaluate the anisotropy of the obtained valence electron density distribution of SrFeO$_3$, a comprehensive comparison using first-principles calculations, considering all electrons, is required. However, in this study, we assume the $3d^4$ and $3d^5$ states to discuss the anisotropy of the localized $3d$ electrons around the Fe site. In the case of $3d^4$, we assume that an electron occupies each $e_g$ orbital with a probability of 1/2. Note that $\rho(\theta, \phi)$ for an isolated ion is related just to the spherical harmonics. A clear anisotropy shows up in the $3d^4$ state in contrast to the completely isotropic electron density in the $3d^5$ state. Here, we extract an approximate relation between the number of Fe $3d$ electrons $N_e$ and $\rho_{\min}/\rho_{\max}$; $N_e = 1.773 \left(\frac{\rho_{\min}}{\rho_{\max}}\right)^2 + 1.163 \frac{\rho_{\min}}{\rho_{\max}} + 2.096$ ($3 \leq N_e \leq 5$) by considering the experimental resolution ($d > 0.28$ Å) (see Supporting Information Section 2 and Figure S4). Since the ratio $\rho_{\min}/\rho_{\max}$ obtained by the CDFS analysis is 0.915, the number of Fe $3d$ electrons is estimated to be $N_e = 4.64(8)$, which is consistent with the previous reports of X-ray absorption spectroscopy measurement ($N_e = 4.7$)[11] and cluster model calculation ($N_e = 4.8$).[26]

Radial distributions of the electron density along the [100] and [111] axes around the Fe site are shown in Figure 3. The electron density in the [111] direction toward the nearest Sr atom has a single-peak structure derived from the $3d$ orbital around $r = 0.2$ Å. The electron density of Fe $3d$ calculated by the Slater-type orbital (STO) of an isolated ion[27] is also plotted as a red broken line, which is in good agreement with the experimental results in the [111] direction. Here, the experimental resolution $d > 0.28$ Å is taken into account when performing the inverse Fourier transform (the details are described in Supporting Information). On the other hand, the electron density in the [100] direction toward the ligand O has one clear peak corresponding to the $3d$ orbital with two shoulder-like structures around $r = 0.5$ and $0.8$ Å. The peak top of the electron density from the Fe nucleus is 0.05 Å closer in the [100] than in the [111] direction. These features may arise from the hybridization between the Fe $3d$ and O $2p$ orbitals.

## 2.3. Valence electron density around the O site

Finally, we focus on the valence electron density around the O site with the site symmetry $4/mm.m$. Since the corresponding valence of Fe obtained by the CDFS analysis was 3.36(8) because of $N_e = 4.64(8)$, the oxygen valence is estimated to be $-1.79(3)$, which deviates from the ideal closed-shell value of $-2$. That is, the valence electron density distribution around the O site should not be isotropic. Figure 4a shows a color map of $\rho(\theta,\phi)$, which is the electron density at a distance $r = 0.40$ Å from the O nucleus, obtained from the CDFS analysis. The observed electron density has some anisotropy. The highest electron density exists toward the surrounding four Sr atoms. On the other hand, the lowest electron density is observed in the [100] direction toward Fe. Brown and green dots in Figure 4b show one-dimensional plots of the valence electron density against the distance from the O nucleus in the [100] and [011] directions, respectively. The difference between the two electron-density profiles is maximum around $r = 0.4$ Å, as shown in the pink line. We also confirmed that there is no significant anisotropy in the radial direction perpendicular to the [100] direction (Figure S6). Blue, orange, and red broken lines show the electron densities of oxygen $2s^2$, $2p^6$, and $2s^22p^6$ calculated by the STO.[27] The experimentally obtained electron density in the [011] direction (green dots) well agrees with to the electron density of oxygen $2s^22p^6$ (red broken line), whereas that in the [100] direction (brown dots) is lower. Furthermore, the difference between the electron densities in the two directions (pink line) mainly corresponds to the contribution of the electron density of oxygen $2p^6$ (orange broken line). These results suggest the existence of ligand holes accommodated in the O $2p_\sigma$—Fe $3d$ antibonding ($\sigma^*$) orbital, which is consistent with previous *p-d* charge-transfer cluster-model calculations[12],[26] and small holes in O $2p$ densities of states predicted by first-principles calculations.[28],[29] Furthermore, while a slight anisotropy in the charge distribution around the Fe site was predicted by the first-principles calculations,[28] the distribution of ligand holes around the O site was captured for the first time by the CDFS analysis.

We observed deviations from ideal $Fe^{4+}$ and $O^{2-}$ states caused by the orbital hybridization as valence electron density distributions. The considerable contribution of the $3d^5\underline{L}$ state seems to be related to the metallic nature[16] and absence of the Jahn-Teller distortion.[23] This unusual state, in which various magnetic phases appear in a magnetic-field,[17],[18] is easily destroyed by slight charge shifts due to oxygen vacancies, resulting in changes in the crystal structure and the physical property.[19]-[24] Thus, the observed *p-d* hybridization may support the unstable charge state of this system. Our experimental technique will be useful in investigating changes in oxidation state due to oxygen vacancies in this system and negative charge transfer states in other systems such as $YBa_2Cu_3O_{7-\delta}$[6] and $SrCoO_3$[9] using the electron density analysis in the future.

## 3. Conclusion

The orbital state in SrFeO$_3$ has been investigated by synchrotron XRD experiments using a high-quality single crystal. We have obtained the valence electron density distribution in SrFeO$_3$ by the CDFS analysis. The number of Fe $3d$ electrons estimated by the anisotropy is $N_e = 4.64(8)$, which indicates the mixed $3d^4$ and $3d^5\underline{L}$ states. The ligand hole is directly observed around the O site. Our experimental observation of the ligand hole may be a touchstone for future theoretical calculations and offers new possibilities for the study of chemical bonding in transition-metal compounds.

## 4. Experimental Section

*Single crystal growth*: Single crystals of SrFeO$_3$ were obtained by high-pressure oxygen annealing of large single crystals of the oxygen-deficient perovskite SrFeO$_{2.5}$ with brownmillerite-type structure as described in Refs.[14],[30] A single crystal of SrFeO$_{2.5}$ was grown by a floating-zone method in an Ar gas flow. The obtained cylindrical crystal with a diameter of about 4 mm was cut into a suitable size for a gold capsule and then treated with oxidizer NaClO$_3$ for 1 hour at 873 K and 8 GPa. Tiny SrFeO$_3$ crystals were obtained by crashing a part of the cylindrical crystal.

*X-ray diffraction measurements*: XRD experiments were performed on BL02B1 at a synchrotron facility SPring-8 in Japan.[31] The dimensions of the SrFeO$_3$ crystal for the XRD experiment were $50 \times 30 \times 30~\mu m^3$. A He-gas-blowing device was employed to cool the crystal to 30 K. The X-ray wavelength was $\lambda = 0.31020$ Å. A two-dimensional detector CdTe PILATUS, which had a dynamic range of ~10$^7$, was used to record the diffraction pattern. The intensities of Bragg reflections with the interplane distance $d > 0.28$ Å were collected by CrysAlisPro program[32] using a fine slice method, in which the data were obtained by dividing the reciprocal lattice space in increments of $\Delta\omega = 0.1°$. Intensities of equivalent reflections were averaged and the structural parameters were refined by using Jana2006.[33]

*Electron density analysis*: A core differential Fourier synthesis (CDFS) method was used to extract the valence electron density distribution around each atomic site.[25],[34] [Kr], [Ar], and [He] type electron configurations were regarded as core electrons for Sr, Fe, and O atoms, respectively. As a result, Sr 5$s$, Fe 3$d$, and O 2$s$/2$p$ valence electrons should remain after the subtraction of the core electron density distribution. The detailed process for extracting the valence electron density distribution is described in Ref. [34].

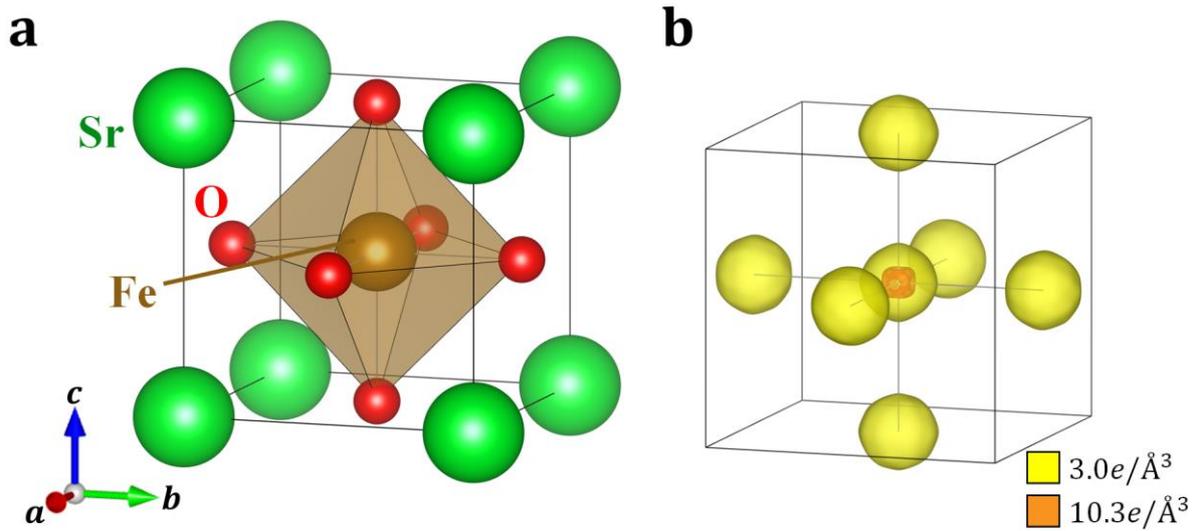

**Figure 1.** (a) Crystal structure and (b) valence electron density distribution of SrFeO$_3$ at 30 K. Yellow and orange iso-density surfaces show electron-density levels of 3.0 and 10.3$e$/Å$^3$, respectively.

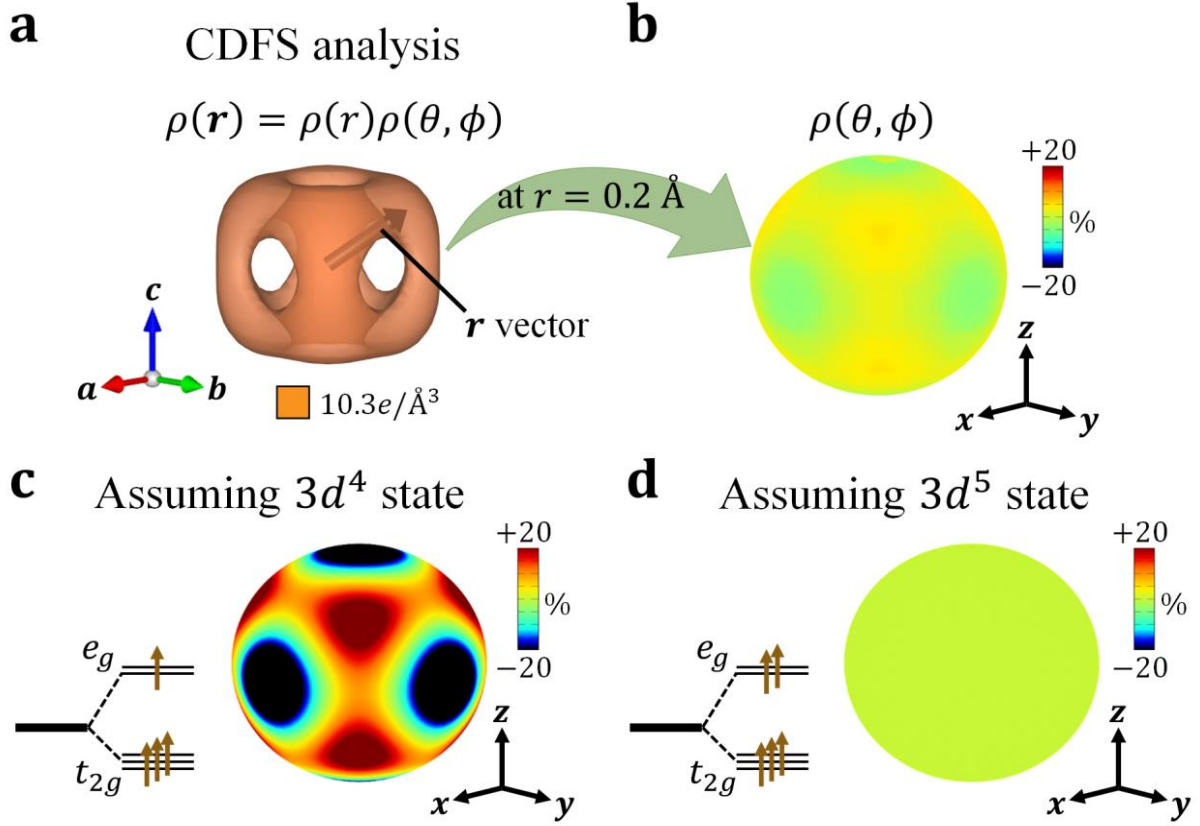

**Figure 2.** (a) Iso-density surface of valence electrons around the Fe site. (b) Color map of the electron density at a distance $r = 0.2$ Å from the Fe nucleus. The $x$-, $y$-, and $z$-axes are parallel to the global $a$-, $b$-, and $c$-axes, respectively. Energy diagrams and color maps of the calculated direction dependence of electron density for the (c) $3d^4$ and (d) $3d^5$ states assuming an isolated Fe atom. The color bar scale is represented by $\frac{[\rho(\theta,\phi) - N_e/4\pi]}{N_e/4\pi} \times 100$ [%]. $N_e$ is the number of $3d$ electrons.

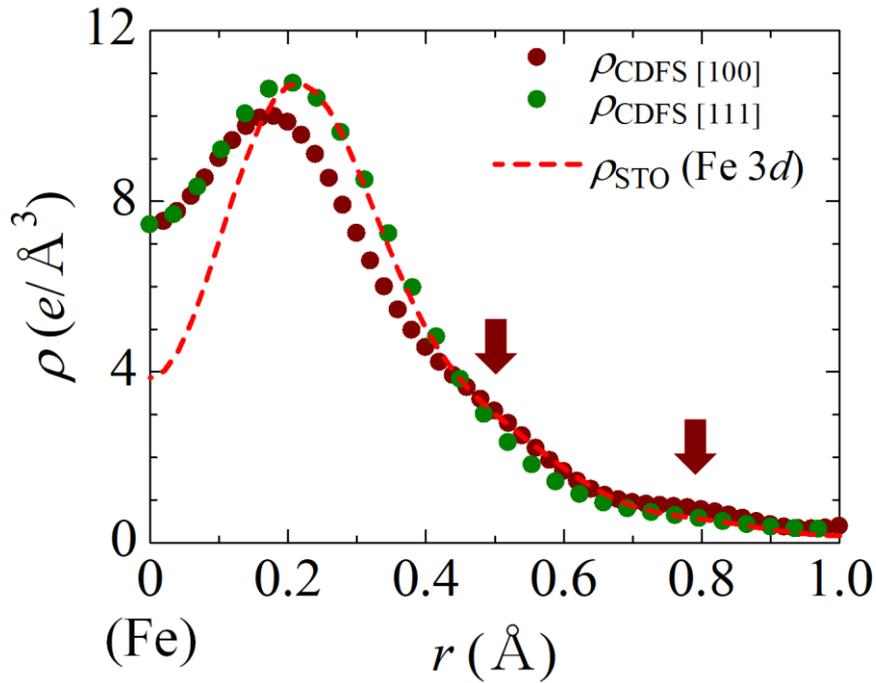

**Figure 3.** Valence electron densities as a function of the distance $r$ from the Fe nucleus. Brown and green dots show the electron densities in the [100] and [111] directions, respectively, obtained by the CDFS analysis. A red broken line shows the $3d$ electron density $\rho_{STO}$ of an isolated Fe ion calculated by the Slater-type orbital[27]. Detailed methods for calculating the valence electron density are described in Supporting Information. Here, the vertical axis for $\rho_{STO}$ is arbitrarily scaled.

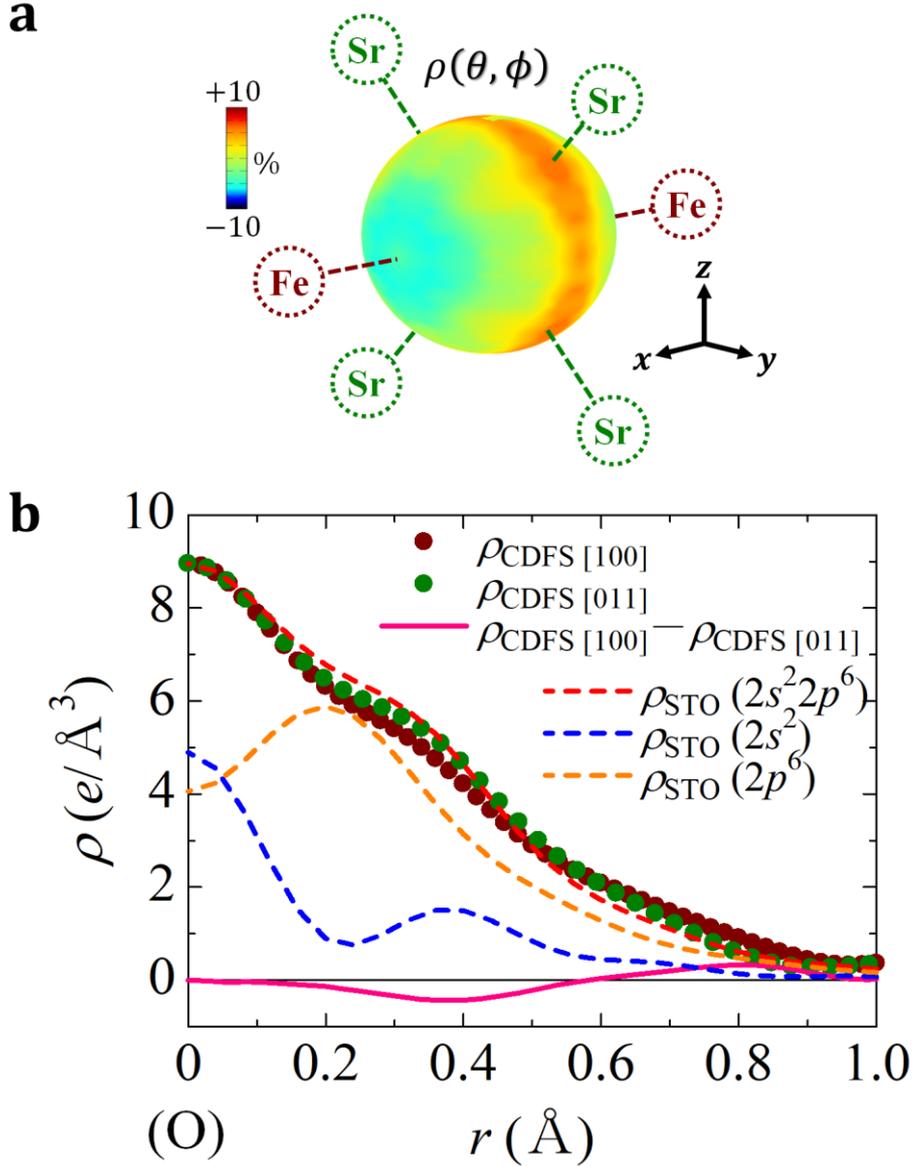

**Figure 4.** (a) Color map of the electron density at a distance $r = 0.4$ Å from the O nucleus at (0, 1/2, 1/2). Fe and Sr atoms are present in the [±100] and [0 ±1 ±1] directions, respectively. (b) Valence electron densities as a function of the distance $r$ from the O nucleus. Brown and green dots show the electron densities in the [100] and [011] directions obtained by the CDFS analysis, respectively. Pink line shows the difference in electron density between the [100] and [011] directions. Blue, orange, and red broken lines show the electron densities $\rho_{STO}$ of oxygen $2s^2$, $2p^6$, and $2s^2 2p^6$ calculated by the Slater-type orbitals[27], respectively. Detailed methods for calculating the valence electron density are described in Supporting Information. The vertical value for $\rho_{STO}$ is arbitrarily scaled.

**Supporting Information**

Supporting Information is available from the Wiley Online Library or from the author.

CCDC 2245476 contains the supplementary crystallographic data for this paper. These data can be obtained free of charge via www.ccdc.cam.ac.uk/data_request/cif, or by emailing data_request@ccdc. cam.ac.uk, or by contacting The Cambridge Crystallographic Data Centre, 12 Union Road, Cambridge CB2 1EZ, UK; fax: +44 1223 336033.

**Acknowledgements**

We thank K. Adachi, and D. Hashizume for in-house XRD characterization of the crystal quality, and H. Sawa, and T. Mizokawa for fruitful discussions. M.G. was a postdoctoral research fellow of the JSPS. This work was supported by a Grants-in-Aid for Scientific Research (No. 20J10988, 22K14010, and 22H00343) from JSPS. The synchrotron radiation experiments were performed at SPring-8 with the approval of the Japan Synchrotron Radiation Research Institute (JASRI) (Proposal No. 2022A1751). Crystal structure and valence electron density distribution are visualized by using VESTA.[35]